\documentclass[useAMS,usenatbib]{mn2e}
\usepackage{graphicx}
\title[KIC\,11911480: the second ZZ\,Ceti in the {\it Kepler} field]{KIC\,11911480: the second ZZ\,Ceti in the {\it Kepler} field}
\author[S. Greiss et al.]{S. Greiss,$^{1}$\thanks{E-mail: s.greiss@warwick.ac.uk} B. T. G\"ansicke,$^{1}$ J.J. Hermes,$^{1}$ D. Steeghs,$^{1,2}$ D. Koester,$^{3}$ G. Ramsay, $^{4}$ \newauthor T. Barclay,$^{5,6}$ D. M. Townsley$^{7}$\\
$^{1}$ Department of Physics, The University of Warwick, CV4 7AL, Coventry, U.K.\\
$^{2}$ Harvard-Smithsonian Center for Astrophysics, 60 Garden Street, Cambridge, MA 02138, USA \\
$^{3}$ Institut f{\"u}r Theoretische Physik und Astrophysik, University of Kiel, 24098 Kiel, Germany \\
$^{4}$ Armagh Observatory, College Hill, Armagh, BT61 9DG, UK\\
$^{5}$ NASA Ames Research Center Institute, Moffett Field, CA 94035, USA \\
$^{6}$ Bay Area Environmental Research Institute, 596 First St West, Sonoma, CA 95476, USA \\
$^{7}$ Department of Physics and Astronomy, University of Alabama, Tuscaloosa, AL, USA}

\begin{document}

\date{}

\pagerange{\pageref{firstpage}--\pageref{lastpage}} \pubyear{2013}

\maketitle

\label{firstpage}

\begin{abstract}
We report the discovery of the second pulsating hydrogen-rich (DA) white dwarf in the {\it Kepler} field, KIC\,11911480. It was selected from the {\it Kepler}-INT Survey (KIS) on the basis of its colours and its variable nature was confirmed using ground-based time-series photometry. An atmosphere model fit to an intermediate-resolution spectrum of KIC\,11911480 places this DA white dwarf close to the blue edge of the empirical boundaries of the ZZ\,Ceti instability strip: $T_\mathrm{eff} = 12\,160 \pm 250 $\,K and $\log{g} = 7.94 \pm 0.10 $. Assuming a mass-radius relation and cooling models for DA white dwarfs, the atmospheric parameters yield: M$_{\rm WD}$ = 0.57 $\pm$ 0.06 M$_\odot$. We also obtained two quarters (Q12 and Q16) of nearly uninterrupted short-cadence {\it Kepler} data on this star. We detect a total of six independent pulsation modes with a $\geq$ 3$\sigma$ confidence in its amplitude power spectrum. These pulsations have periods ranging between 172.9 s and 324.5 s, typical of the hotter ZZ Ceti stars. Our preliminary asteroseismic study suggest that KIC\,11911480 has a rotation rate of 3.5$\pm$0.5 days. 
\end{abstract}


\section{Introduction}

Most stars end their lives as white dwarfs (WDs), and hence studying the galactic population of WDs offers insight into the star formation history of the Galaxy. WDs are very diverse and can be found in single or binary systems, and their study is hence central to a global understanding of stellar evolution. 

Around 80\% of WDs have hydrogen-dominated atmospheres, also known as DA WDs, making them the most commonly found and studied class of WDs \citep{koesteretal79, shipman79, bergeronetal92, giammicheleetal12, kleinmanetal13}. As WDs cool, they pass through instability strips, exhibiting periodic variations in their mean intensity with amplitudes of up to a few percent. Four classes of pulsating WDs are known, depending on atmospheric composition: the hot pre-WDs (PG\,1159 or DOV stars), warm helium-atmosphere WDs (V777\,Her or DBV stars, $T_\mathrm{eff} \simeq22\,000-29\,000$\,K), cool hydrogen-atmosphere WDs (ZZ\,Ceti or DAV stars, $T_\mathrm{eff} \simeq10\,900-12\,300$\,K), and the carbon-atmosphere WDs (DQV, $T_\mathrm{eff}\simeq20000$\,K, \citealt{dufouretal08-2, montgomeryetal08}). The latter class is the most recently discovered, and there is still a question as to if the variability in DQVs is caused by pulsations or magnetic spots \citep{dunlapetal13, lawrieetal13}. Pulsating DA WDs have the appropriate temperatures to foster a hydrogen partial ionization zone, which in turn drives global non-radial $g$-mode pulsations \citep{robinsonetal82,wingetetal91}.\\

While WD parameters are traditionally inferred from model fits to spectra, precision asteroseismology of WDs has the tantalising potential to probe the masses and compositions of their electron-degenerate cores, as well as their non-degenerate envelopes \citep[e.g.][]{winget+kepler08-1, fontaine+brassard08-1}, to determine their internal rotation profiles \citep{charpinetetal09-1}, to measure weak magnetic fields \citep{wingetetal91}, to search for planetary companions via pulse timing variations \citep{mullallyetal08-1}, and to constrain nuclear reaction rates (e.g. ${}^{12}\mathrm{C}(\alpha,\gamma)^{16}\mathrm{O}$, \citealt{metcalfeetal02-1}). \\
 
Only a handful of WDs were known in the NASA \textit{Kepler} field \citep{boruckietal10} when it was launched in 2009, and none of them exhibited pulsations \citep{ostensenetal10-1, ostensenetal11-2}. The recent discoveries of the first V777\,Her star (KIC\,8626021, \citealt{ostensenetal11-2}) and the first ZZ\,Ceti star (KIC\,4552982, \citealt{hermesetal11}) within the \textit{Kepler} field represents the dawn of precision space-based WD asteroseismology. ZZ\,Cetis are known to have pulsation periods ranging from 100\,s to 1000\,s, which can vary from one end of the instability strip to the other \citep[e.g.][]{mukadametal04-1}, although pulsating extremely low-mass (M$<0.25 \mathrm{M_\odot}$) WDs have pulsation periods ranging from $1184-6235$~s \citep{Hermes13}. The amplitudes of individual pulsation modes in cool ZZ\,Cetis can vary dramatically on time scales of days to weeks \citep[e.g. G29-38,][]{kleinmanetal94-1,  kleinmanetal98-1}.  This presents a challenge for finding modes with closely-spaced frequencies, such as rotationally split modes, because the beating between such modes appears as an amplitude variation of a single mode in short (e.g. 1 night) time series.  This ambiguity can be partially ameliorated with repeated observations from a single site, though gaps between consecutive nights leads to mode aliases.  In order to remove ambiguity altogether, or to study amplitude variation itself, an actual continuous time series is essential. The Whole Earth Telescope \citep{natheretal90} has achieved nearly continuous series for time spans up to several days to weeks for some objects with carefully coordinated ground-based campaigns \citep{wingetetal91, wingetetal94, dolezetal06, provencaletal12}. The quarter-long, continuous, uniformly sampled time series provided by {\it Kepler} present a novel opportunity to affirm methods for working around aliasing ambiguity in ground-based studies.

It is clear that a full understanding of WD structure and evolution will require a larger sample of targets. With this goal in mind, we began the search for ZZ\,Ceti stars in the {\it Kepler} field down to $r$ = 19.5 mag. Here we present the detection and study of the second pulsating DA WD in the {\it Kepler} field: KIC\,11911480. We begin in Section \ref{selection} by describing our selection method and show its optical spectrum and fit in Section \ref{spec}. In Section \ref{astero}, we focus on the asteroseismic analysis of the light curve of KIC\,11911480, obtained from nearly six months of short-cadence observations on the {\it Kepler} spacecraft spread over 15 months. Finally, we present our conclusions in Section \ref{conclusion}.

\section{Target selection}
\label{selection}

\begin{figure}
\noindent
\parbox{\columnwidth}{
\centering
\hspace*{-.5cm}
\includegraphics[trim=0cm 0cm 0cm 0cm, clip, angle=0, scale=1]{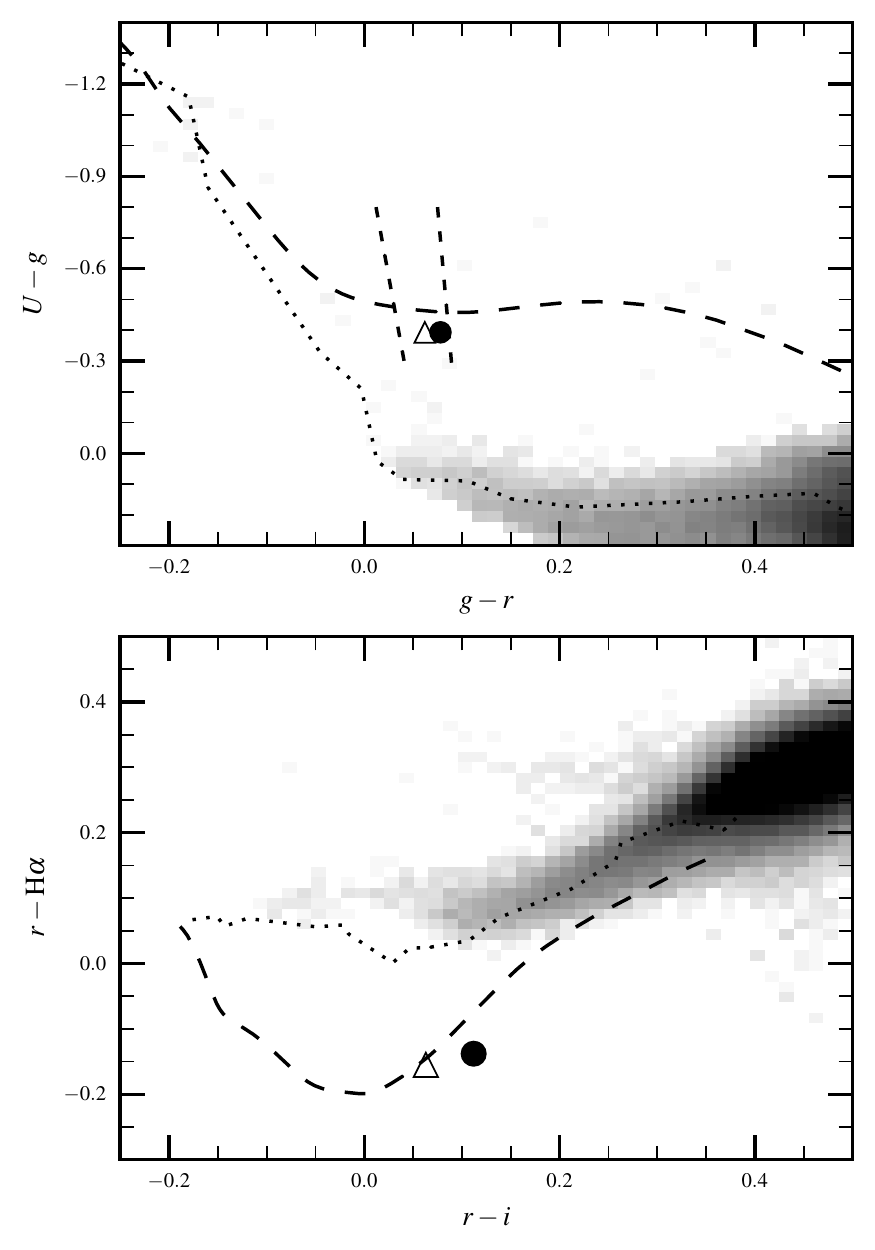}}
\smallskip
\caption{\small $(U-g, g-r)$ and ($r-{\mathrm H}\alpha, r-i$) colour-colour diagrams of stellar sources from the first half of the \textit{Kepler}-INT survey (gray scale), with $\log{g} = 8$ DA WD cooling tracks (dashed line). The dotted line indicates the Pickles main sequence tracks taken from \citet{grootetal09}. The filled circle corresponds to KIC\,4552982 \citep{hermesetal11}, whereas the open triangle marks KIC\,11911480. They are clearly blue and H$\alpha$ deficient objects. The vertical dashed lines in the upper panel correspond to the empirical boundaries of the ZZ\,Ceti instability strip \citep{gianninasetal11} projected in the $(U-g, g-r)$ colour space. \label{ccd}}
\end{figure}

We began our selection of ZZ\,Ceti candidates using ($U-g, g-r$) colours from the {\it Kepler}-INT Survey (\citealt{greissetal12}, see upper panel Fig. \ref{ccd}). KIS is a deep optical survey of the {\it Kepler} field, using four broadband filters, $U, g, r, i$ and one narrowband filter, H$\alpha$. All the observations are taken using the Wide Field Camera (WFC) on the 2.5m Isaac Newton Telescope (INT) on the island of La Palma. We narrowed down our selection to candidates in, or close to, the empirical ($T_{\rm eff}$, $\log{g}$) instability strip \citep{gianninasetal11} projected into ($U-g, g-r$) space using the cooling models presented in \cite{tremblay_and_bergeron09}. We also make use of the H$\alpha$ filter to confidently select DA WDs, as they have broad H$\alpha$ absorption lines and therefore stand out as H$\alpha$-deficient object in the ($r-{\rm H}\alpha, r-i$) colour-colour diagram (lower panel of Fig.\,\ref{ccd}). \\

In our photometric selection, we found a number of candidates including KIC\,4552982, the ZZ\,Ceti star discovered by \citet{hermesetal11}. Another star, KIC\,11911480, showed significant variability from ground-based observations obtained as part of the RATS-{\it Kepler} survey \citep{ramsayetal13} which consists of one hour sequences of 20\,s $g$-band exposures of objects in the {\it Kepler} field using the INT/WFC. The power spectrum of RATS-{\it Kepler} data of KIC\,11911480 revealed short-period variability with a dominant signal at $\sim$290\,s. Table\,\ref{info} gives the coordinates and magnitudes of KIC\,4552982 and KIC\,11911480. \\

\begin{table}
\caption{Coordinates and KIS magnitudes of KIC\,11911480 and KIC\,4552982. \label{info}}
\begin{center}
\begin{tabular}{l c c}
\hline \hline
 & KIC\,11911480 & KIC\,4552982 \\
 \hline
RA (J2000) & 19:20:24.90 & 19:16:43.83 \\
Dec (J2000) & +50:17:21.3 & +39:38:49.7 \\
$U \dotfill$ & 17.701 $\pm$ 0.022 & 17.362 $\pm$ 0.007 \\
$g \dotfill$ & 18.094 $\pm$ 0.015 & 17.755 $\pm$ 0.005 \\
$r \dotfill$ & 18.032 $\pm$ 0.021 & 17.677 $\pm$ 0.007 \\
$i \dotfill$ & 17.969 $\pm$ 0.033 & 17.565 $\pm$ 0.009 \\
H$\alpha \dotfill$ & 18.187 $\pm$ 0.059 & 17.815 $\pm$ 0.018 \\
K$_p \dotfill$ & 17.63 $\pm$ 0.05 & 17.85 $\pm$ 0.05\\
\hline
\end{tabular}
\\
\end{center}
K$_p$ corresponds to the magnitude from the Kepler Input Catalog (KIC, \citealt{brownetal11})
\end{table}

\section{Optical Spectroscopy}
\label{spec}

\begin{figure}
\noindent
\parbox{\columnwidth}{
\centering
\hspace*{-.5cm}
\includegraphics[angle=-90,scale=0.325]{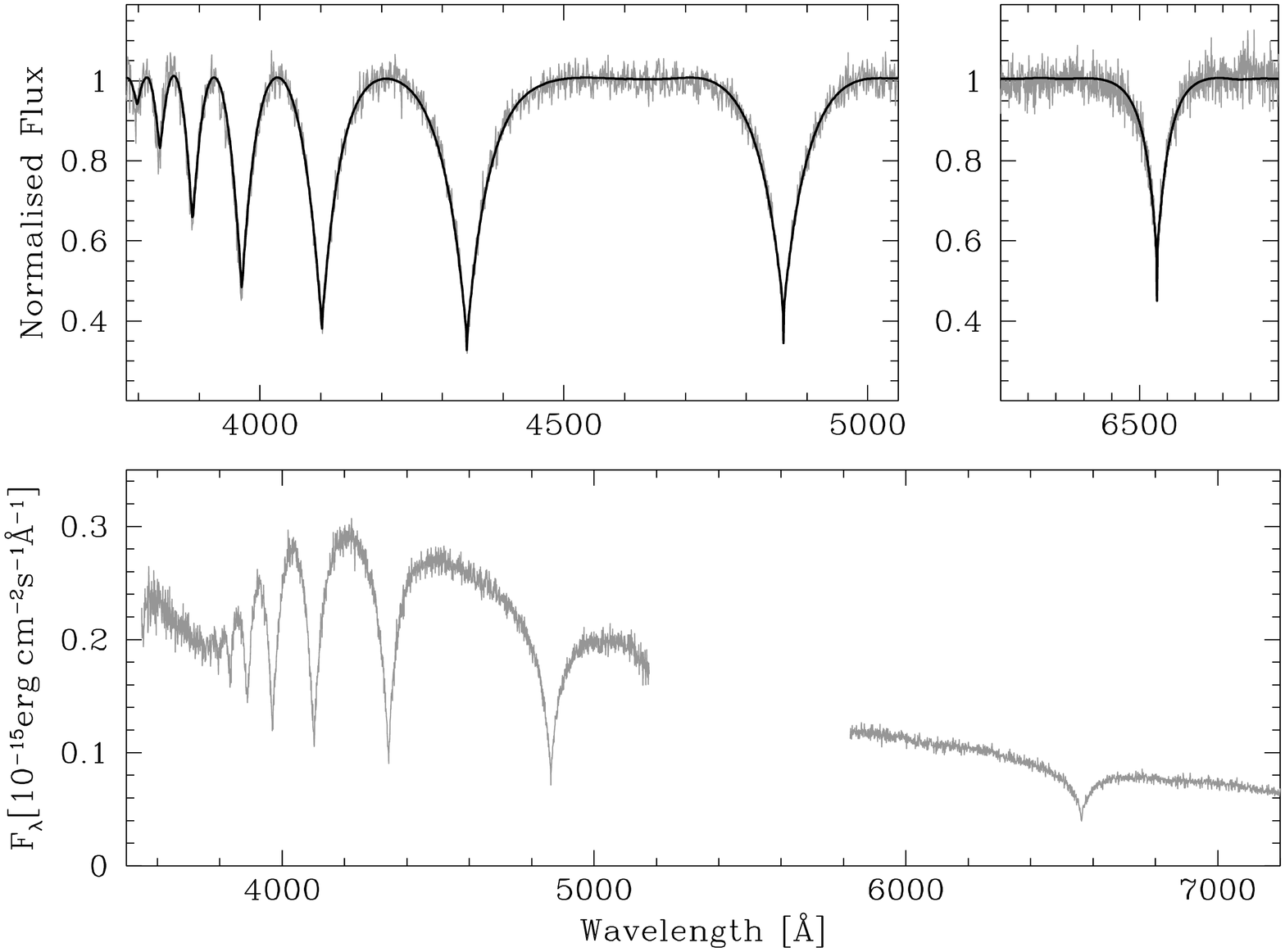}}
\smallskip
\caption{\small Bottom panel: flux-calibrated WHT spectrum of KIC11911480 plotted in grey. Top panel: normalised Balmer profiles were used to determine its atmospheric parameters using 1-D atmospheric models: $T_\mathrm{eff} = 12\,350 \pm 250 $\,K and $\log{g} = 7.96 \pm 0.10 $. \label{spectrum}}
\end{figure}

On June 7$^{th}$ 2013, we obtained spectroscopy of KIC\,11911480 using the double-armed Intermediate Resolution Spectrograph\footnote{http://www.ing.iac.es/Astronomy/instruments/isis/} (ISIS) on the William Herschel Telescope (WHT) on the island of La Palma. We observed under 1" seeing conditions. We used the R600R and R600B gratings, in the ISIS blue and red arms respectively, with a 1" slit. 
The blue arm was centred at 4351$\mathrm{\AA}$ and the red arm at 6562$\mathrm{\AA}$. The blue spectra covered a total wavelength range from $\sim$3700$\mathrm{\AA}$ to $\sim$5000$\mathrm{\AA}$, and the red spectra ranged from $\sim$5700$\mathrm{\AA}$ to $\sim$7200$\mathrm{\AA}$ (see Fig.\,\ref{spectrum}). The resolution of the spectrum in the red arm is $\sim$2\,${\rm \AA}$ and $\sim$1.8\,${\rm \AA}$ in the blue arm. Three consecutive 20 minute exposures were taken in order to increase the signal-to-noise ratio (S/N) of the average spectrum.

The spectra were de-biased and flat-fielded using the standard STARLINK\footnote{The STARLINK Software Group homepage website is http://starlink.jach.hawaii.edu/starlink.} packages KAPPA, FIGARO and CONVERT. Optimal spectral reduction was then done with PAMELA\footnote{PAMELA was written by T. R. Marsh and can be found in the STARLINK distribution ‘Hawaiki’ and later releases.} \citep{marsh89}. Copper-argon arc lamp exposures were taken at the start and end of each night for the wavelength calibration of the spectra. We identified around 10 to 15 arc lines in each arm, which we fitted with fourth order polynomials. We observed two standard stars for the flux calibration of the spectra: Wolf\,1346 and Grw+70 5824. We used MOLLY\footnote{MOLLY was written by T. R. Marsh and is available from http://www.warwick.ac.uk/go/trmarsh/software.} for the wavelength and flux calibration of the extracted 1-D spectra.

We calculate the average spectrum of KIC\,11911480 using the three spectra obtained from the individual 20 minute exposures and obtain a S/N $\sim$ 33 at 4400$\mathrm{\AA}$. The normalised Balmer line profiles in the average spectrum were then fitted using the WD models of \cite{koester10}, following the procedure described in \cite{homeieretal98}. For the computation of the models, we used the Stark-broadened Balmer line profiles of \cite{tremblay_and_bergeron09} and adopted the ML2/$\alpha = 0.8$ prescription for convection. In model atmospheres, convection has been described with the mixing length theory \citep{bohm-vitense58}, and ML2/$\alpha$ ($\alpha$ is the mixing length to pressure scale height ratio) is a parametrisation typically used for WD atmospheres \citep{tassouletal90}. \\

In Fig.\,\ref{spectrum}, we show the spectrum and overplot the best model fit, which returned the following parameters for the DA WD: $T_\mathrm{eff} = 12\,350 \pm 250 $\,K and $\log{g} = 7.96 \pm 0.10 $. The uncertainties in the atmospheric parameters were estimated from fitting the three individual 20\,min WHT spectra and taking the root mean square. This places KIC\,11911480 close to the blue edge of the empirical boundaries of the ZZ\,Ceti instability strip \citep{gianninasetal11}. Using a mass-radius relation and the evolutionary cooling models from \cite{fontaineetal01} with a carbon-oxygen core \cite{bergeronetal01}\footnote{The cooling models can be found on http://www.astro.umontreal.ca/$\sim$bergeron/CoolingModels/. Also refer to \cite{holberg+bergeron06-1, tremblayetal11b, bergeronetal11} for colour and model calculations.}, we obtain a mass estimate of our ZZ\,Ceti star: M$_{\rm WD}$ = 0.58 $\pm$ 0.06 M$_\odot$. However, we note that this may be a slight overestimate of the true WD mass, as it is now well established that the spectroscopically determined surface gravities, and hence the masses, of WDs with temperatures $\la13,000$\,K, are systematically too high (see e.g. \citealt{bergeronetal90, gianninasetal11, koesteretal09b}, for a discussion). The most likely explanation for this problem is that 1-D mixing length theory does not properly account for the effects of convection on the temperature structure of the atmosphere \citep{tremblayetal11a}. For completeness, we adopt the 3-D models corrections of \cite{tremblayetal13} and find $T_\mathrm{eff} = 12\,160 \pm 250 $\,K and $\log{g} = 7.94 \pm 0.10$ corresponding to M$_{\rm WD}$ = 0.57 $\pm$ 0.06~M$_\odot$. The differences in our case are negligibly small and we adopt the corrected values for this study.

\begin{figure}
\centerline{\includegraphics[angle=270,width=\columnwidth, scale=1.2]{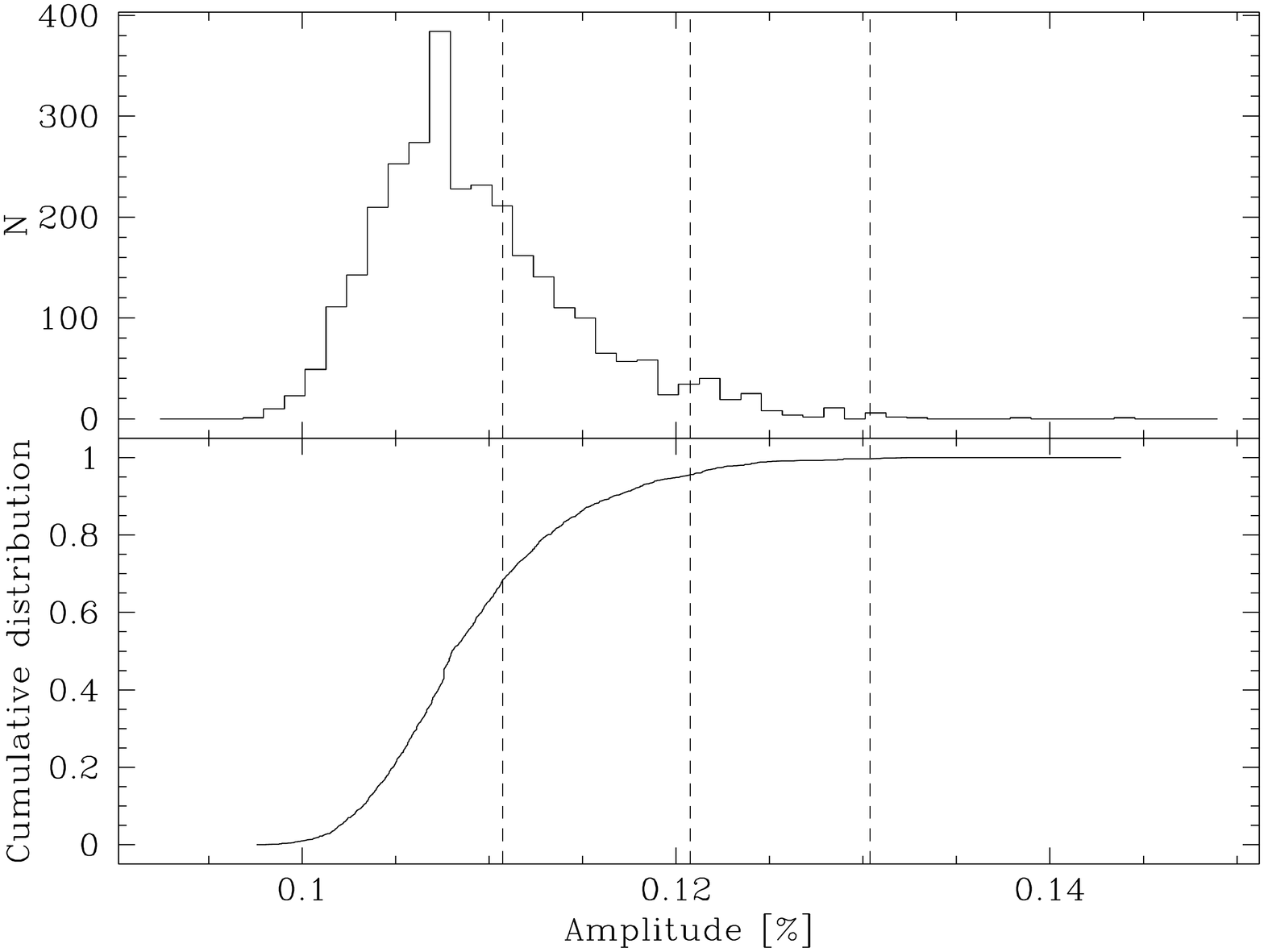}}
\caption{\label{f-mc_sigma} Top panel: Distribution of the highest
  amplitude recorded in 3000 randomised light curves. Bottom panel:
  Cumulative distribution of the highest amplitudes. The 68.3\%,
  95.5\% and 99.7\% detection thresholds are indicated by the dashed
  vertical lines.}
\end{figure}

\section{Preliminary asteroseismic study}
\label{astero}

After confirmation of its pulsating nature, we were awarded {\it Kepler} short-cadence mode observations of KIC\,11911480 during Quarters 12 and 16 (Q12, Q16). {\it Kepler} operates in two modes: short and long cadence observations consisting of 58.89 s and 29.4 min exposures respectively. The Q12 observations were taken between January 4$^{th}$ and March 28$^{th}$ 2012 and Q16 spanned from January 11$^{th}$ to April 8 $^{th}$ 2013. Each {\it Kepler} quarter corresponds to three months of observations. However, during Q12, a series of coronal mass ejections affected the run by bringing the duty cycle of Q12 down to 88.2\%. During Q16, {\it Kepler} went into rest mode for 11.3 days, leading to a duty cycle of 84.8\% for that quarter. 
The {\it Kepler} light curves, produced using simple aperture photometry and delivered by the {\it Kepler} Science Operations Center pipeline, were downloaded from the MAST website\footnote{http://archive.stsci.edu/kepler/publiclightcurves.html}. Details on the data handling, processing and releases can be found on the {\it Kepler} data analysis website\footnote{http://keplerscience.arc.nasa.gov/DataAnalysis.shtml}.

We calculate the discrete Fourier Transforms (FT) of each set separately, using the TSA package within MIDAS written by Alex Schwarzenberg-Czerny. 
We observe optical variability in the {\it Kepler} data with periods ranging from 137.1\,s to 519.6\,s (see Table\,\ref{frequencies-table} for more details). These periods are within the expected values for ZZ\,Ceti stars, which are known to have $g$-mode pulsations ranging from 100 to 1000\,s \citep{fontaine+brassard08-1}, and match the pulsation periods of other known hot ZZ\,Ceti stars \citep{mukadametal04-1}.

\begin{table*}
\caption{Pulsation frequencies of KIC\,11911480 from the Q12 and Q16 data. The uncertainties are given in between brackets. $\Delta f$ corresponds to the frequency spacing between two consecutive frequencies in the table\label{frequencies-table}}
\begin{tabular}{l l l l l l l l l}
\hline
& P (P$_{\rm err}$) (s) & $f$ $(f_{\rm err})$ $(\mu\rm{Hz})$ & A (\%) & $\Delta f (\mu\rm{Hz})$ & P (P$_{\rm err}$) (s) & $f$ $(f_{\rm err})$ $(\mu\rm{Hz})$ & A (\%) & $\Delta f (\mu\rm{Hz})$\\
 \hline
 \smallskip
 &  Q12 &  &  & Q16 &  &  \\
$f_{1,-}$ & 290.9664 (7) & 3436.823 (8) & 0.185  &  & 290.9675 (3) & 3436.810 (4) & 0.439 & \\
$f_{1,o}$ & 290.8016 (1) & 3438.770 (1) & 1.187  & 1.947 & 290.8026 (6) & 3438.759 (7) &  2.175 &1.949 \\
$f_{1,+}$ & 290.6322 (4) & 3440.775 (5) & 0.368  & 2.005 & 290.6341 (2) & 3440.753 (3) & 0.641 & 1.994 \\
\\
$f_{2,-}$ & 259.3731 (2) & 3855.451 (3) & 0.501 & & 259.3738 (1) & 3855.440 (2) & 0.997 & \\
$f_{2,o}$ & 259.2531 (2) & 3857.235 (3) & 0.581 & 1.784 & 259.2538 (1) & 3857.224 (2) & 0.975 & 1.784 \\
$f_{2,+}$ &  -  & - & - & - & 259.1352 (3) & 3858.989 (4) & 0.391 & 1.764\\
\\
$f_{3,-}$ & 324.529 (1) & 3081.39 (1) & 0.169 & & 324.5299 (6) & 3081.381 (5) & 0.321 & \\
$f_{3,o}$ & 324.3152 (9) & 3083.420 (9) & 0.185 & 2.03 & 324.3175 (6) & 3083.398 (6) & 0.278 & 2.017 \\
$f_{3,+}$ & 324.1032 (1) & 3085.44 (2) & 0.100 $\star$ & 2.02 & - & - & - & - \\
\\
$f_{4,-}$ & 172.9588 (4) & 5781.72 (1) & 0.086 $\star$ &  & - & - & - & -\\
$f_{4,o}$ & 172.9003 (5) & 5783.68 (1) & 0.113 $\star$ & 1.96 & 172.9015 (3) & 5783.64 (1) & 0.149 & \\
\\
$f_{5,o}$ &  - & - & - & - & 202.5687 (6) & 4936.60 (1) & 0.118  $\star$ & \\
$f_{5,+}$ &  - & - & - & - & 202.4873 (5) & 4938.58 (1) & 0.091 $\star$ & 1.98\\
\\
$2\,f_{1,o}$ & 145.4007 (3) & 6877.54 (1) & 0.125 & & 145.4013 (1) & 6877.516 (6) & 0.301 & \\
\\
$f_{1,o}+f_{2,o}$ & 137.0610 (2) & 7296.02 (1) & 0.158 &  & 137.0614 (1) & 7295.999 (7) & 0.234 &  \\
\\
$f_{4,o}-f_{1,o}$ & 426.936 (3) & 2342.27 (2) & 0.085 $\star$ & & 426.455 (2) & 2344.91 (1) & 0.140 & \\
\\
$f_{4,-}-f_{2,-}$ & 519.600 (4) & 1924.56 (2) & 0.078 $\star$ & & - & - & - & -\\
$f_{4,o}-f_{2,o}$ & 519.110 (4) & 1926.37 (1) & 0.078 $\star$ & 1.81 & 519.093 (4) & 1926.440 (1) & 0.119 $\star$ & \\
$f_{4,+}-f_{2,+}$ &  - & - & - & - & 518.636 (2) & 1928.136 (7) & 0.088 $\star$ & 1.696 \\
\hline
\end{tabular}
\smallskip
\\
$\star$ the frequency was detected below 3$\sigma$
\end{table*}

\subsection{Significance threshold}

We adopted a randomisation technique to determine the detection
thresholds for the {\it Kepler} light curve. In this process, we bootstrap the
{\it Kepler} data, keeping the times of the individual observations in
place, but randomise the sequence of the fluxes. For each of these randomised light curves we compute the discrete FT as described above and we record the highest amplitude from each of the randomisation. The process of randomising is repeated a large number of times, building up a smooth cumulative distribution of the highest recorded amplitudes (Fig.\,\ref{f-mc_sigma}), from which the 3$\sigma$ threshold is determined, i.e. 99.7\% of the highest amplitudes recorded in the randomised light
curves fall below that threshold. We found that the 3$\sigma$ threshold converges after about a thousand randomisations, and, to err on the side of caution, we used 3000 randomisations for the final calculation of the significance thresholds. In the case of each individual quarter, the 3\,$\sigma$ thresholds for Q12 is 0.138\% and Q16 are 0.130\%. Strictly speaking, this threshold only applies to the highest amplitude signal detected in the {\it Kepler} light curves, and the entire process would need to be
repeated after pre-whitening the {\it Kepler} data with the highest
amplitude signal, carrying on in an iterative fashion until no signal
satisfies the 3\,$\sigma$ threshold. However, we have experimented
with a {\it Kepler} light curve that had all pulsation signals (Table\,\ref{frequencies-table})
and spurious signals removed, and the resulting 3\,$\sigma$ threshold
is 0.132\% in the case of the Q12 data. It is therefore not significantly different from the one derived from the original {\it Kepler} light curve. 

Note that this method yields more conservative significant thresholds than the more widely adopted method which uses 4$\langle {\rm A}\rangle$ as the limit to consider significant peaks, where $\langle {\rm A}\rangle$ corresponds to the average amplitude of the amplitude power spectrum (see Section 5.4 of \citealt{aertsetal10} and references therein). We find that 4$\langle {\rm A}\rangle$ = 0.113\% for the Q12 run, and 4$\langle {\rm A}\rangle$= 0.099\% for Q16.

\begin{figure*}
\noindent
\centering
\hspace*{-0.75cm}
\includegraphics[scale=1]{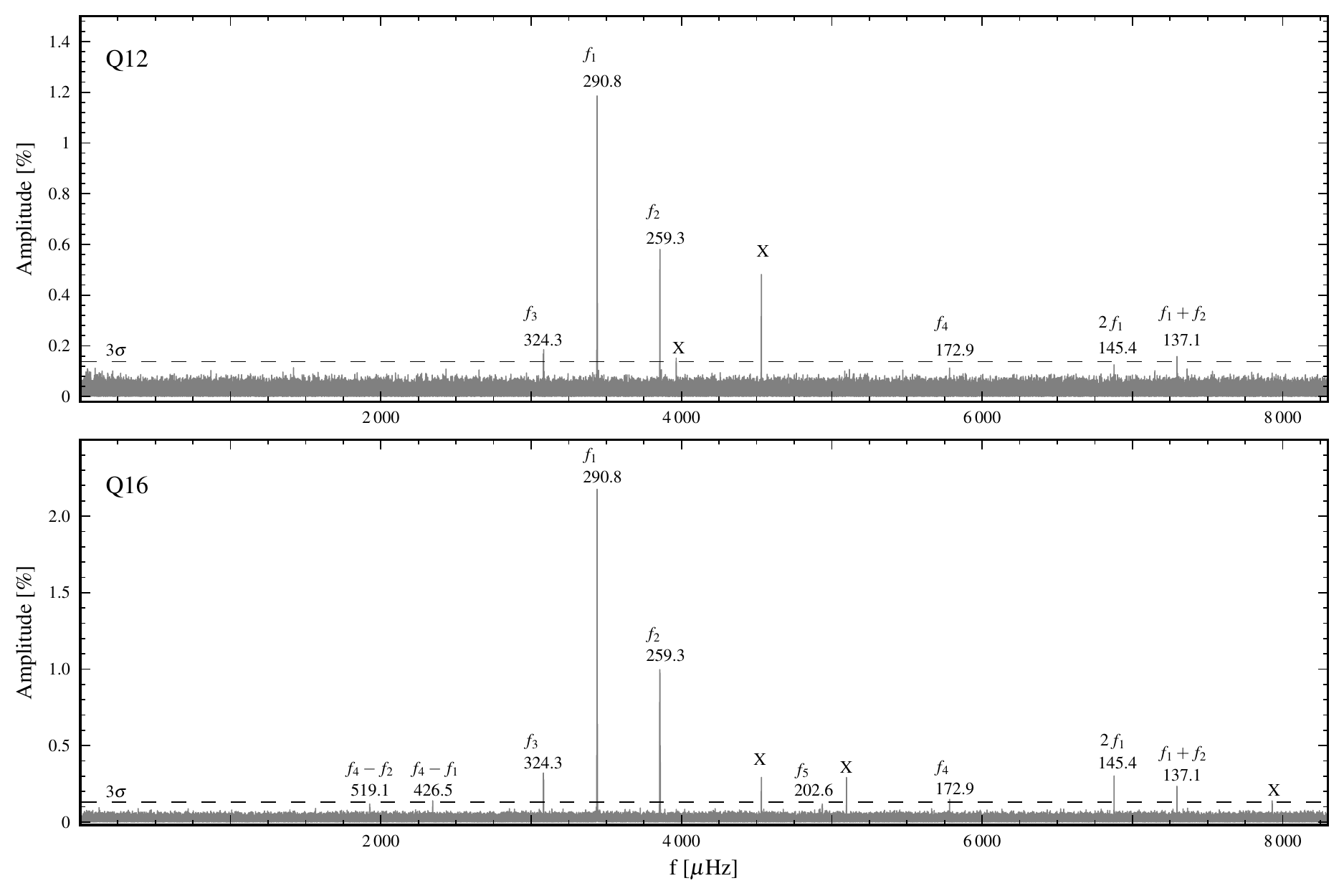}
\smallskip
\caption{\small Amplitude power spectrum of KIC\,11911480 from the Q12 (top) and Q16 (bottom) data. The Xs above certain frequencies indicate the {\it Kepler} spurious frequencies and the dashed horizontal line corresponds to the 3\,$\sigma$ threshold limit. The values noted above each significant frequency corresponds to period (in seconds) of the given pulsation mode. \label{power-spectrum-all}}
\end{figure*}

\subsection{Pulsation modes}

\begin{figure}
\hspace*{-0.5cm}
\vspace*{-0.5cm}
\includegraphics[trim=0cm 0cm 0cm 0cm, clip, scale=0.9]{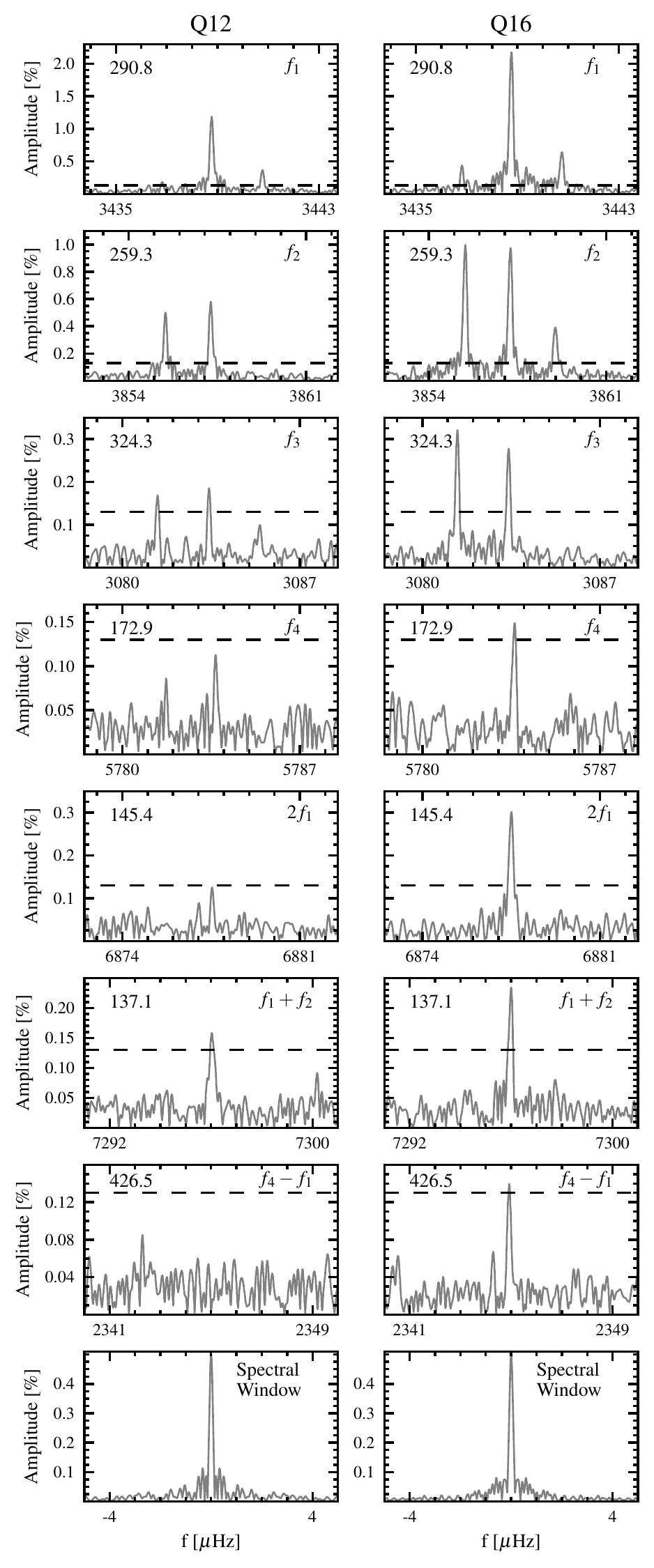}
\smallskip
\caption{\small {\it Kepler} amplitude power spectra of KIC\,11911480, our first ZZ\,Ceti discovered in the {\it Kepler} field from KIS. The panels on the left correspond to the Q12 data, whereas the ones on the right correspond to the Q16 data. The dashed lines correspond to the 3$\sigma$ significance threshold in each dataset. The top left hand-side of each panel shows the corresponding period (in seconds). The bottom panels in both column show the spectral window of each quarter. Splitting of the modes is a direct indication of the star's rotation (note that it is common that not all modes of a multiplet are detected at a particular epoch, see e.g. Table 5 of Kepler et al. 2003). \label{power-spectrum}}
\end{figure}

In the amplitude power spectra of KIC\,11911480 (Fig.\,\ref{power-spectrum-all}), we mark the {\it Kepler} spurious frequencies taken from the Data Characteristic handbook\footnote{The Data Characteristic handbook can be downloaded from the following webpage: http://keplerscience.arc.nasa.gov/DataAnalysis.shtml}. These spurious frequencies are multiples of 1/LC period, where LC corresponds to long-cadence exposures of 29.4 min. Additional spurious frequencies, unrelated to the inverse of the LC period, have been detected from the analysis of many early {\it Kepler} quarters. Some components in the power spectrum of KIC\,11911480 appear and disappear between both quarters and it is clear that the amplitudes of the detected frequencies are generally stronger in Q16 than in Q12 (see Fig.\,\ref{power-spectrum}). The error on the amplitudes is 0.028\%. This large overall amplitude variation between Q12 and Q16 may be explained by the fact that our target was observed with two different custom masks in both quarters.

We calculate the frequencies and amplitudes of each significant peak, using a least-squares sine wave fitting routine at each individually selected peak from the FT. We find seven pulsation modes detected above the 3\,$\sigma$ threshold in the Q16 data set, four of which are independent (Table\,\ref{frequencies-table}). In the Q12 data, we find five pulsations modes above the 3\,$\sigma$ threshold, out of which three are independent. The pulsation modes found in Q12 are all detected in Q16 as well. In Table\,\ref{frequencies-table}, we also add two pulsation periods, $f_{5}$ and $f_{4}-f_{1}$, which are not significantly detected in Q12 but they are very close to the Q16 3\,$\sigma$ threshold. Their amplitudes are larger than the 4$\langle {\rm A}\rangle$~=~0.099\% for Q16. The main reason why we believe the detection of $f_{5}$ is because it shows splitting with the same frequency separation as the other significant modes. Also, $f_{4}-f_{1}$ is a non-linear combination of two significant frequencies. In total, we find that KIC\,11911480 has five independent pulsation modes and four combination frequencies.  Non-linear combination frequencies are not generated by the same physical mechanism driving the pulsations of the star. \cite{brickhill92} showed that these combination frequencies may come from the distortion of the sinusoidal waves associated to the normal modes travelling from the convective to the radiative zone of the star, where the heat transport changes dramatically at the base of the hydrogen ionization zone (see also \citealt{wu+goldreich99, vuille00, wu01, yeatesetal05}). Their amplitudes can provide information on the physical conditions in the WD convection zone \citep{montgomery05}. 

Also, we look into the phases of these modes and find that they are coherent enough to produce an O-C diagram over the 15-month {\it Kepler} observations but are indicative of a large drift in phase. We will address the analysis of those phase changes in a future paper as this requires a careful treatment of the Q12 and Q16 data obtained with different pixel masks.

We also notice splitting of some of the modes, which are denoted with `+' or `-' signs and placed next to the central component (`o') of each pulsation mode in Table\,\ref{frequencies-table}. Not all components of a multiplet are always detected (see Fig.\,\ref{power-spectrum}). A full asteroseismic study is beyong the scope of this paper, yet we have attempted to match the observed periods to adiabatic pulsation models with the constraints provided by our spectroscopic mass and temperature determinations. The models of \citet{Romero12} of a 0.57 M$_\odot$, $12\,101$ K WD with a thick ($10^{-3.82}$ M$_{\rm WD}$) hydrogen layer mass are in decent agreement with the observed periods of $f_1-f_4$ if these four modes have $\ell=1$ and $k= 4, 3, 5, 2$, respectively. However, this is only qualitative guess at a solution, and a full asteroseismic analysis is required to arrive at a more secure identification of these modes.

\subsection{Rotation rate}

We see what appears to be multiplet splitting of some modes, which is a direct manifestation of the star's rotation rate (Fig.\,\ref{power-spectrum}). In the limit of slow rotation, the difference between the frequency of one mode of indices $l, k, m$ ($\sigma_{k,lm}$) and the frequency in the non-rotating case ($\sigma_{k,l}$) is:
\begin{equation}
\sigma_{k,l,m} - \sigma_{k,l} = m (1 - C_{k,l}) \Omega
\end{equation}
where $C_{k,l}$ comes from the Coriolis force term in the momentum equation and $\Omega$ is the rotation frequency \citep{wingetetal91, vauclair97}. Note that this equation is the classical first order expansion. In the asymptotic limit for $g$-modes, $C_{k,l}$ only depends on the degree of the mode: $C_{k,l} \simeq \frac{1}{l (l+1)}$. When a pulsating WD rotates, each mode of degree $l$ can be split into 2$l$+1 components. We see splitting into three components in several modes in the power spectrum of KIC\,11911480 (see Fig.\,\ref{power-spectrum}), which likely corresponds to an $\ell=1$ mode in those cases, leading to $C_{k,l} \simeq 0.5$. The frequency spacing between the split components of the modes is quite consistent, $1.93\pm0.10 \mu\rm{Hz}$, suggesting these modes are all of the same spherical degree. This corresponds to a rotation rate of $3.0\pm0.2$ days. However, $f_1-f_4$ (with periods from $172.9-324.5$ s) are likely low-radial-order and far from the asymptotic limit, so their $C_{k,l}$ values should not be identical, and are not exactly 0.5. If we adopt the $C_{k,l}$ values of the model from \citet{Romero12} discussed in Section 4.2, we obtain a rotation rate of $3.5\pm0.2$ days. To best reflect the systematic uncertainties, we adopt a rotation rate of $3.5\pm0.5$ days.

Notably, the small but significant deviations in the observed frequency splittings provide additional asteroseismic information, especially useful for constraining which modes are trapped by composition transition zones \citep{Brassard92}. The shorter-period g-modes have lower radial order, and these splittings are observed to have values of 1.97~$\mu\rm{Hz}$ for $f_1$, 1.77~$\mu\rm{Hz}$ for $f_2$, 2.03~$\mu\rm{Hz}$ for $f_3$, and 1.94~$\mu\rm{Hz}$ for $f_4$.

This value is in agreement with previous rotation frequencies found in ZZ\,Ceti stars. \cite{fontaine+brassard08-1} give an overview on pulsating WDs and provide the asteroseismic rotation rates of seven ZZ\,Ceti stars, spanning from 9 to 55 hours, i.e. 0.4 to 2.3 days. In the case of non-pulsating WDs, the sharp NLTE core of the H$\alpha$ line in their spectra has been used in many studies to measure the projected rotation velocities of the stars \citep{heberetal97, koesteretal98, karletal05}. In all cases, the same conclusion was drawn: isolated WDs are generally {\it slow rotators}.

\section{Conclusion}
\label{conclusion}

We report on the discovery of the second ZZ\,Ceti in the {\it Kepler} field: KIC\,11911480. It was discovered using colour selections from the {\it Kepler}-INT Survey and confirmed with ground-based time-series photometry from the RATS-{\it Kepler} survey. Follow-up {\it Kepler} short-cadence observations during Q12 and Q16 are analysed: five independent pulsation modes, as well as four non-linear combinations, were detected in the combined power spectrum of KIC\,11911480, all ranging from 137.1\,s to 519.6\,s. The splitting of four of the independent pulsations enable us to estimate the rotation period of the star to be $3.5\pm0.5$ days, assuming these are all $\ell=1$ modes.

An intermediate-resolution spectrum using ISIS on the WHT and DA WD model atmosphere fits returned our ZZ\,Ceti's atmospheric parameters: $T_\mathrm{eff} = 12\,160 \pm 250 $\,K and $\log{g} = 7.94 \pm 0.10 $. This places KIC\,11911480 close to the blue edge of the empirical boundaries of the ZZ\,Cetis instability strip. Using DA WD cooling models and evolutionary tracks, the surface gravity translates to M$_{\rm WD}$~=~0.57~$\pm$~0.06~M$_\odot$. \\

{\it Kepler} has concluded observations of this hot ZZ\,Ceti star, amassing more than 210,000 frames of this pulsating WD. We encourage follow-up, ground-based observations, which will allow to measure the amplitude of any nonlinear combination frequencies with periods $<120$ s, which fall below the {\it Kepler} Nyquist frequency. Additionally, ground-based data can extend the time baseline in order to measure the rate of period change of the pulsation modes excited in this ZZ\,Ceti star.

\section*{Acknowledgments}
SG acknowledges support through a Warwick Postgraduate Research Scholarship. 
The research leading to these results has received funding from the European Research Council under the European Union's Seventh Framework Programme (FP/2007-2013) / ERC Grant Agreement n. 320964 (WDTracer). BTG was supported in part by the UK’s Science and Technology Facilities Council (ST/I001719/1). DS acknowledges support from STFC through an Advanced Fellowship (PP/D005914/1) as well as grant ST/I001719/1.

This paper makes use of data collected at the Isaac Newton Telescope and the William Herschel Telescope, operated on the island of La Palma, by the Isaac Newton Group in the Spanish Observatorio del Roque de los Muchachos. 

Balmer/Lyman lines in the models were calculated with the modified Stark broadening profiles of \cite{tremblay_and_bergeron09} kindly made available by the authors.

This paper also includes data collected by the {\it Kepler} mission. Funding for the {\it Kepler} mission is provided by the NASA Science Mission directorate. The {\it Kepler} data presented in this paper were obtained from the Mikulski Archive for Space Telescopes (MAST). STScI is operated by the Association of Universities for Research in Astronomy, Inc., under NASA contract NAS5-26555. Support for MAST for non-HST data is provided by the NASA Office of Space Science via grant NNX09AF08G and by other grants and contracts.

\label{lastpage}

\end{document}